\documentclass[prl,letterpaper,twocolumn,showpacs,superscriptaddress]{revtex4}
\usepackage{graphicx,psfrag,amsmath,amssymb,amsfonts,bbm,latexsym,color,dcolumn}

\begin{document}

\title{Microscopic origin of low frequency flux noise in Josephson circuits}
\author{Lara Faoro}
\author{Lev B. Ioffe}
\affiliation{Department of Physics and Astronomy, Rutgers University, 136 Frelinghuysen
Rd, Piscataway 08854, New Jersey, USA}
\date{\today }

\begin{abstract}
We analyze the data and discuss their implications for the microscopic origin
of the low frequency flux noise in superconducting circuits. We argue that
this noise is produced by spins at superconductor insulator boundary whose
dynamics is due to RKKY interaction. We show that this mechanism explains
size independence of the noise, different frequency dependences of the spectra 
reported in large and small SQUIDs and gives the correct 
intensity for realistic parameters.
\end{abstract}

\pacs{85.25.Cp, 03.65.Yz,73.23.-b}
\maketitle



Studies of the flux noise in superconducting structures have a long history
that began in 80s with the demonstration that it is the flux and not the
critical current noise that limits the sensitivity of dc SQUIDs
(Superconducting QUantum Interference Devices) \cite{Koch1983,Wellstood1987}%
. The noise phenomenological characterization performed at this time revealed
many puzzling features that defied a simple model of its microscopic origin,
so this problem was put aside and largely forgotten. Recently, the interest
to it was renewed when it was realized that the flux noise limits the
coherence of qubits based on superconducting devices \cite{Devoret2004} and
that the dephasing of 'flux' qubits is due by low frequency flux noise with
intensity comparable to the one measured in dc-SQUIDs \cite%
{Yoshihara2006,Kakuyanagi2007}. It is likely that the same flux noise will
limit the quantum coherence in 'phase' qubits \cite{Martinis2002}.

Recently two models for the excess $1/f$ flux noise were proposed. The first
one\cite{Koch2007} proposes that the flux noise is due to the electrons that
hop between traps in which their spins have fixed, random orientations. The
second model \cite{DeSousa2007} attributes the noise to the electrons that
experience spin flips induced by the interaction with tunneling Two Level
Systems (TLSs) and phonons. Both models rely on some assumptions that are
difficult to justify: for example, in order to match the intensity of the
noise spectral density reported in the experiments, the number of \emph{%
thermally} activated TLSs present in the oxide layer has to be much larger
than in a typical glass at the same temperature. For instance, in a typical
loop of radius $R=1\mu m$ and volume $10^{7}$nm$^{3}$ the observed noise
value implies activation of $10^{5}-10^{6}$ spin fluctuators with magnetic
moment $\mu _{B}$ while a typical glass of the volume has about $10$
thermally excited TLSs at $T=0.1\text{K}$. \ This Letter has two goals: (i) to
present a critical analysis of the flux noise phenomenology and its
implication for the possible models of its microscopic origin, (ii) to
propose a novel mechanism in which the low frequency noise is due to spin
diffusion on the superconductor surface generated by the exchange mediated
by the conduction electrons. We demonstrate that this spin dynamics together
with the spatial dependence of the surface current density on the thin
superconducting SQUID loop leads to low frequency $1/f^{\alpha }$ flux noise
spectral density with ${\alpha \in \{0,1\}}$ and that the intensity of the
noise does not depend on the area of the SQUID, as long the ratio $R/W$
remains constant ($W$ denotes the width of the SQUID line); however, the
details of diffusion in large SQUIDs, with ${W\sim 100\mu m}$ and small
SQUIDs having ${W\sim 1\mu m}$ might be different. In particular, the
frequency dependence of the noise spectrum might vary depending on the size
of the SQUIDs and the measured frequency range.

All experiments agree on the magnitude of the noise at frequency ${f\sim 1%
\text{Hz}}$ and its area independence. Specifically, Wellstood \cite%
{Wellstood1987} observed noise spectra ${S_{\Phi }^{1/2}(1\text{Hz})\approx
(4-10)\mu \Phi _{0}\text{Hz}^{-1/2}}$ at temperature below ${0.1\text{K}}$
in $\text{Nb}$ and $\text{Pb }$dc SQUIDs with sizes in the range ${R\,,W\sim
(30-300)\mu \text{m}}$ on $\text{Si/SiO}$ substrate with $\text{ Nb/NbO/PnIn}
$ Josephson junctions. Cromar et al \cite{Cromar1989} reported the value ${%
S_{\Phi }^{1/2}(1\text{Hz})=2.3\mu \Phi _{0}\text{Hz}^{-1/2}}$ for $\text{Nb}
$ SQUIDs device with very high quality $\text{Nb/AlO/Nb}$ junctions at ${4%
\text{K}}$. Finally, Bialczak et al \cite{Bialczak07} measured ${S_{\Phi
}^{1/2}(1\text{Hz})=2\mu \Phi _{0}Hz^{-1/2}}$ at ${20\text{mK}}$ in a Al
SQUID loop of size ${W\sim 1\mu \text{m }}$ on sapphire substrate with $%
\text{Al/AlO/Al}$ Josephson junctions. Although the details of temperature
dependence were studied only in \cite{Wellstood1987}, all experiments agree
that the noise does not decrease at very low temperatures. Similarly, all
data show homogeneous noise spectra where single strong fluctuators cannot
be resolved. The frequency dependence of the noise is more controversial.
Namely, Wellstood's flux noise power spectra in the frequency range ${%
(1-10^{3})\text{Hz}}$ $\ $displayed ${1/f^{\alpha }}$ dependence with
exponent ${\alpha =0.66}$ at low temperatures ${0.022 \text{K}<T<1\text{K}}$ and with
exponent ${\alpha =1}$ at ${1\text{K}<T<4.2\text{K}}$. The data \cite{Cromar1989} show the
dependence $f^{-0.7}$ in the interval $(400-$ $10^{3})$ Hz at ${4\text{K}}$, below $%
400$ Hz the frequency dependence decreases to approximately $f^{-0.1}$ and
completely ceases in the $(0.1-40)$ Hz interval. Bialczak et al reported $%
1/f^{\alpha }$ spectrum with $\alpha =0.95$ in the frequency range ${%
(10^{-5}-1)\text{Hz at }20\text{mK}}$. The temperature dependence of the
noise \cite{Wellstood1987} shows two different temperature regimes: at ${<0.5\text{K}}$ the noise is $T$ independent, while at ${1\text{K}<%
T<4.2\text{K}}$ it displays $T^{2}$ dependence with the crossover
regime (${0.5\text{K}<T<1\text{K) }}$ that is non-monotonous in some
samples. Two distinct regimes suggests two different microscopic mechanisms
for the noise at low and high temperature; in the following we shall focus
on the low temperature regime. As it will be clear below, a very important
piece of information is the high frequency cutoff of the $1/f^{\alpha }$
dependence. Unfortunately, no direct measurements are available but the
observed dephasing of the flux qubits indicates that this cutoff is at least 
$10\text{MHz}$\cite{Nakamura2006}.

We now discuss the implications of the data for the noise origin. The noise
persistence at low temperatures indicates that it is due to a subsystem
characterized by very low energy scales, smaller than the minimal
temperature available experimentally ($\sim 20\text{mK}$). This rules out the thermally
excited TLSs \cite{Koch2007} or vortices and points towards weakly
interacting nuclear or electron spins. The spin mechanisms agrees also with
the observation of homogeneous low frequency noise power spectra (which is
incompatible with the vortex origin). Nuclear spins can be excluded for
three reasons: first, the flux produced by each spin scales as $1/L$ (where $%
{L\sim R,W}$ is the linear size of the device) thereby leading to ${S_{\Phi
}^{1/2}(\omega )\propto L^{1/2}}$ while data are roughly size independent;
second, \emph{all} frequency scales associated with nuclear spins are very
low ($f<1\text{kHz}$) in contrast with the results of the dephasing analysis which
shows that $1/f$ persists up to $10\text{MHz}$ \cite{Nakamura2006}; third,
one expects that nuclear spin noise would be substrate dependent\cite%
{Koch2007}. Paramagnetic electron spins located on the superconductor or
insulator interfaces seem to be more promising candidates since their
contribution to the flux noise is roughly size independent. Properties of
these spins were extensively studied for $\text{Si/SiO}_{2}$ interfaces. ESR
experiments have shown that: (i) the surface density of spins varies between 
${\nu _{2D}\approx 10^{10}-10^{12}\text{cm}^{-2}}$ \cite{Schenkel2006}; (ii)
the $g$ factor of these spins is isotropic and it has value ${g=2.00136\pm
0.00003}$ \cite{Stesmans2000}. As we show below, such surface density is
barely sufficient to explain the level of flux noise at ${1\text{Hz}}$ \emph{%
if} one assumes that \emph{all} these spins remain active at low
temperatures but is difficult to reconcile them with schemes in which only a
small percentage of the spins remain active at low $T$ \cite{Koch2007}. The
value $g\cong 2$ shows that the spin orbit coupling is very weak indicating
that the interaction between paramagnetic spins and TLSs is very small in
contrast to assumptions of Refs. \cite{Koch2007},\cite{DeSousa2007}.

The dynamics of the electron spins in the insulator substrate is due to the
interaction with other electron spins or with surrounding nuclei. For a
dilute spin system such as $\text{Si/SiO}_{2}$ interfaces all energy scales
associated with these interactions are too small to account for the wide
frequency range observed experimentally: both dipole-dipole interaction
between electron spins with density ${\nu _{2D}\approx 10^{12}\text{cm}^{-2}}
$ and their interaction with nuclear moments of Si that have natural
concentration of $5\%$ correspond to ${f\approx 10\text{kHz}}$. Estimating
the total flux noise produced by these spins, i.e. $\int S(\omega )d\omega $
we get ${\sim \left( \mu _{0}\mu _{B}\right) ^{2}(R/W){\nu _{2D}\Phi _{0}^{2}%
}}$ with ${\left( \mu _{0}\mu _{B}\right) ^{2}\sim 10^{-26}\text{cm}^{2}}$
which is of the right order of magnitude but somewhat smaller than the
observed noise. We conclude that these spins in the insulator are unlikely
to provide the dominant source of noise.

The energy scales are much larger for the electron spins in the proximity of
the superconductor which allows RKKY interaction between them which is a
much stronger coupling than dipole-dipole interaction in the insulator. This
interaction is due to the virtual scattering of conduction electrons off a
magnetic impurity described by the Kondo Hamiltonian: $\displaystyle{H_{K}=%
\mathcal{J}\hat{S}\cdot \hat{\sigma}}$ where ${\hat{S}}$ is the spin
operator for the impurity, $\hat{\sigma}$ is the spin operator of a
conduction electron and $\mathcal{J}$ is the exchange constant. Integrating
out the conduction electrons one gets: 
\begin{equation}
H_{RKKY}=\sum_{i,j}V(r_{ij})\hat{S}_{i}\hat{S}_{j}\;  \label{RKKYe}
\end{equation}%
where ${\displaystyle{V(r)=V_{0}(r)e}}^{-2r/\xi }{r^{-3}\cos \varphi }$ and $%
\varphi $ changes quickly on the length scale of the Fermi wavelength ${%
\lambda _{F}}$ \cite{Vavilov2003}. The interaction 'constant' $V_{0}(r)$ is
a weak function of the distance, it is controlled by the electron density of
states $\nu $ and Kondo temperature:\ $\displaystyle{V_{0}(r)=(2\pi
)^{-1}\nu \mathcal{J}^{2}(r)}$, with $\displaystyle{\mathcal{J}(r)=2[\nu \ln
^{2}(v_{F}/(rT_{K}))]^{-1}}$ so that the average interaction at $r\ll \xi $ 
\begin{equation}
\left\langle V^{2}(r)\right\rangle ^{1/2}=\frac{{1}}{2\sqrt{2}\pi \nu r^{3}}%
\left( \frac{2}{\ln [v_{F}/(rT_{K})]}\right) ^{2}\;  \label{RKKYint}
\end{equation}%
Because of this interaction, the magnetization $M(t,r)$ of spins
averaged over the volume that contains ${N\gg 1}$ spins obeys the diffusion
equation: 
\begin{equation}
\left[ \frac{d}{dt}-{\mathcal{D}}\nabla ^{2}\right] M(t,r)=0
\label{DiffusionEquation}
\end{equation}%
with diffusion coefficient ${\mathcal{D}}$ which depends on the typical
distance between the spins on the surface, i.e ${r=\sqrt{\nu _{2D}}\approx
(10-10^{2})~\text{nm}}$ and the average interaction $\left\langle
V^{2}(r)\right\rangle ^{1/2}$ (\ref{RKKYe}). Typical electron density of
states for Al, Pb and Nb are respectively: ${\nu _{\text{Al}}=35/eV\text{nm}%
^{3}}$, ${\nu _{\text{Pb}}=44/eV\text{nm}^{3}}$ and ${\nu _{\text{Nb}}=160/eV%
\text{nm}^{3}}$. Assuming Kondo temperatures ${T_{K}\approx 0.01-1\text{K}}$, we
estimate: 
\begin{equation}
\mathcal{D}=r^{2}\left\langle V^{2}(r)\right\rangle ^{1/2}\approx
(10^{8}-10^{9})~~\text{nm}^{2}\text{s}^{-1}  \label{diff}
\end{equation}%
This model neglects a few important physical effects. First, it neglects the
spin orbit scattering and assumes that the diffusion process involves only
electron spins located on the SI interface. As a result, the total
magnetization $M$ of the spins in contact with the superconductor is
conserved. Second, the estimate for ${\mathcal{D}}$ Eq.(\ref{diff}) assumes
that the spins are in direct contact with the metal. However, paramagnetic
spins responsible for the flux noise are likely to be located in the surface
oxide of the thickness ${d=(2-3)\text{nm }}$ with some of them further away
from the superconducting wire. For impurity located at depth ${y}$ from the
superconductor the strength of RKKY interaction decreases as $\displaystyle{%
V(r,y)\sim e^{-2y/a_{0}}V(r)}$, where $a_{0}$ is the atomic distance. A more
realistic model should include 'fast' spins at the surface with the
diffusion constant given in Eq. (\ref{diff}) and slower spins coupled to the
'fast' subsystem by a weakened RKKY interaction. Third, the diffusion
approximation for the spin dynamics neglects the effect of the rare pairs of
spins located at distances much smaller than the average distance between
the spins. Such spins are strongly coupled with each other, the difference
in the energy of their triplet and singlet state is much larger than their
coupling to their neighbors, so they change their state rarely. This
mechanism generates an additional noise at low frequencies.

In order to find the effective flux $\Phi_{eff}$ produced by the spin magnetization we
determine the spin energy $E$ in the field of the test current $I$ in the loop. We find that:
\begin{equation}
\Phi_{eff}=\frac{d E}{d I}=  g \mu_B \int \frac{\hat{S}(r) B(r)}{I} d^2r \;
\label{fluxeff}
\end{equation}
Here $\mu _{B}$ is the Bohr magneton, $\hat{S}(r)$ is the spin density operator and $B(r)$ denotes the probing magnetic field. Conservation of the total magnetization by
spin diffusion means that it would not produce any noise if the probing
magnetic field were uniform. In fact, it is not: the SQUID loop is
typically a strip conductor of width $W$ greater than its thickness with
length $L\gg W$. For the film thickness less or comparable with the
penetration depth $\lambda $, the dependence of the current density on $x$
near the center of the strip is $\displaystyle{J_{s}(x)=2 I/(\pi
W)[1-(2x/W)^{2}]^{-1/2}}$ for ${-W/2+\lambda <x<W/2+\lambda }$, while the
current density falls away exponentially to zero at the edges ${\pm W/2}$ 
\cite{SDC}. This current density results in a probing magnetic field $\displaystyle{B(x)=\frac{\mu
_{0}}{2}J_{s}(x)}$.
 The spin diffusion together with the divergency of the surface
current density close to the edges of the loop generates $1/f$ flux spectral
density: 
\begin{equation}
\langle \Phi_\tau \Phi_0 \rangle =\left ( g \mu _{B} \right )^{2} L \int_{-W/2}^{W/2}dxdx^{\prime } \frac{\hat{S}_\tau (x)B(x)\hat{S}_{0}(x^{\prime })B(x^{\prime })}{I^2}\;  
\label{corr}
\end{equation}
where $x$ is the coordinate across the wire strip. To
compute the integral in Eq. (\ref{corr}) we expand the spin density operator $\hat{S}_{t}(x)
$ as a series of orthonormal eigenfunctions of the diffusion equation (\ref%
{DiffusionEquation}) with boundary conditions ensuring zero magnetization
current at the wire boundary, i.e. $\displaystyle{\frac{d}{dt}M_{t}(x=\pm
W/2)=0}$: 
\begin{equation}
\hat{S}_{t}(x)=\sqrt{\frac{2}{W}} \sum_{q=\pi n/W}\hat{S}_{q}(t)\cos \left[ \left( x+\frac{W}{2}\right)
q\right] \;  \label{four}
\end{equation}%
where $n$ is positive integer.
By substituting the expansion given in Eq. (\ref{four}) into Eq. (\ref{corr}%
) we find: 
\begin{equation}
\langle \Phi_\tau \Phi_0 \rangle=(g \mu _{B})^{2} L \sum_{q}B_{q}^{2}\langle \hat{S}_{q}(\tau )\hat{S}_{q}(0)\rangle \;
\label{corrq}
\end{equation}
where
\begin{eqnarray} 
B_{q}&=&\sqrt{\frac{2}{W}}\int_{-W/2}^{W/2}dx \frac{B(x)}{I} \cos \left[ \left( x+\frac{%
W}{2}\right) q\right] \;\notag \\
&=& \frac{\mu_0}{\sqrt{2 W}} {\mathcal J}_0 \left (\frac{q W}{2} \right ) \cos \left(\frac{q W}{2} \right ) \;
\label{bq}
\end{eqnarray}
where ${{\mathcal J}_{0}(x)}$ is the Bessel function. The Fourier transform of the spin density correlator (\ref{corrq}) is found from the solution
of the diffusion equation (\ref{DiffusionEquation}): 
\begin{equation}
\langle \hat{S}_q^2(\omega) \rangle=\frac{\sigma_s}{2} \frac{{\mathcal{D}}q^{2}}{%
\omega ^{2}+({\mathcal{D}}q^{2})^{2}}\;  \label{spettro}
\end{equation}
where ${\sigma_s=\rho d}$ is the surface density of paramagnetic spins $\frac{1}{2}$.
We can define two frequency regimes: small frequencies with ${f \ll f_W}$ and large frequencies with ${f \gg f_{W}}$. ${f_{W}}$ is the characteristic equilibrium frequency for spins that
diffuse across SQUID of width $W$: 
\begin{equation}
f _{W}=\frac{\mathcal{D}}{W^{2}}=\left\{ 
\begin{array}{ll}
10^{-2}-10^{-1}\text{Hz} & ~~\mbox{if $W \sim 100 \mu$m}; \\ 
10^{2}-10^{3}\text{Hz} & ~~\mbox{if $W \sim 1 \mu$m}.%
\end{array}%
\right.   \notag
\end{equation}%
At small frequencies, the flux noise spectrum given in Eq. (\ref{corrq}) is white, with noise
amplitude given by: 
\begin{equation}
\langle \Phi^2\rangle_{\omega \to 0} = \left (\frac{\mu_0 \mu_B}{2 \pi}  \right )^2 \sigma_s \frac{L}{W}
 \frac{{\mathcal J}_0(\pi)^2}{f_W} \,;  \label{spettro0}
\end{equation}%
where ${{\mathcal J}_{0}(\pi )=-0.3042}$. 
At large frequencies, Eq. (\ref{bq}) reduces to $\displaystyle{B_q= \sqrt{\frac{2}{\pi}} \frac{\mu_{0}}{W} 
\frac{1}{\sqrt{q}}}$
and the flux noise spectrum given in Eq.(\ref{corrq}) becomes: 
\begin{eqnarray}
\langle \Phi^2 \rangle&=&\frac{2}{\pi^2} (\mu _{0} \mu_B)^{2} \sigma_s \frac{L}{W} \int_{0}^{\infty }\frac{dq}{q}\frac{{\mathcal{D}}q^{2}}{\omega ^{2}+({\mathcal{D}}q^{2})^{2}} \; \notag \\
&=& \frac{4}{\pi }\left( \mu _{0}\mu _{B}\right)
^{2}\sigma_s \frac{R}{W}\frac{1}{f }\;  \label{spettrof}
\end{eqnarray}
where we have written explicitly the length of the SQUID loop ${L=2 \pi R}$.
At intermediate frequencies, we expect a crossover between $1/f$ and white
noise behavior. It is quite straightforward to estimate the intensity of the $1/f$
noise. 
Assuming that ${\sigma_s \approx 10^{16}m^{-2}}$ (similar density was reported
for $\text{Si/SiO}_{2}$ interfaces) and that ${R/W\sim 10}$ we find flux
noise spectral density ${S_{\Phi }(1\text{Hz}) \approx 3(\mu \Phi _{0})^{2}%
\text{Hz}^{-1}}$, in agreement with the observed "universal" value. Because
the noise is due to the spins on the surface, its level has the same ${R/W}$
dependence as in Ref \cite{Bialczak07}.

Thus, the spin diffusion model explains the excess flux noise measured in
large SQUIDs \cite{Wellstood1987},\cite{Cromar1989}, which spectra
correspond to the intermediate/high frequencies regime, but not the $1/f$
noise observed in much smaller devices \cite{Bialczak07} since the latter
was measured in the range corresponding to the low frequency regime where
purely spin diffusion model predicts a constant spectral density. However,
two physical effects missing in the model that were mentioned above are very
likely to produce a significant low frequency noise in the smaller SQUIDs:
the presence of weakly coupled spins further away from superconductor and
the presence of strongly coupled spin pairs. Indeed, assuming flat
distribution of the spin depth inside the insulating layer one gets an
exponential distribution of the coupling to the spins on SI interface ${P({%
\mathcal{J}})\propto 1/{\mathcal{J}}}$ that directly translates into the $%
1/f $ spectrum of the noise generated by these spins. The intensity of this
noise is determined by the areal density, $\nu _{2D}^{\prime }$ of the spins
in the layer of approximately atomic depth ${\sim 2a_{0}}$. Generally, one
expects $\nu _{2D}^{\prime }\sim \nu _{2D}$; a significant difference in the
values of $\nu _{2D}^{\prime },\nu _{2D}$ might lead to a complicated
frequency dependence of the noise observed in a wide frequency range: e.g. a 
$1/f$ behavior at high frequencies (due to 'fast' spin diffusion) followed
by a partial saturation at intermediate frequencies which is followed again
by $1/f$ regime at very low frequencies due to deep and weakly coupled
spins. Our preliminary analysis shows that close pairs of spins strongly
coupled to each other by RKKY interaction also lead to $1/f$ contribution to
the low frequency noise.

Finally, we discuss experimental tests of the proposed model. The crucial
ingredient of our analysis is the rough temperature independence of the noise below $200$ mK \cite{Wellstood1987}, it would be important to verify it
for small devices. The spin origin of the noise can be tested by applying a
significant external magnetic field. If this field is larger than the local
field, $B_{\text{loc}},$ produced by the spin neighbors, the spin rotates
around the axis determined by the external field. If it is orthogonal to the
probing field, fast rotation of the spin implies that the effective spin
noise is shifted to high frequencies. If these fields are parallel, the
effect is much less. The effective probing field acting on the spins on the
insulator boundary inside the SQUID loop is mostly perpendicular to the
surface of the sample, while the probing field of the spins on SI boundary
is parallel to it. Thus, applying magnetic field in different directions one
can verify the spin mechanisms and determine the spin location. The local
field that should be exceeded in these experiments is of the order of $B_{%
\text{loc}}\lesssim 100G$ for the spins $10\text{nm}$ apart on SI surface
and of the order of $B_{\text{loc}}\lesssim 0.1\text{G}$ for the spins in
the insulator. Random position of spins in these models implies that there
will be always strongly coupled spin pairs capable of producing \ the low
frequency noise but the number of such pairs should go down rapidly with
field. The validity of the spin models discussed in this paper can be also
tested directly by fabrication of the samples with decreased density of spin
defects on the surface of the insulator and by protecting the surface of
superconductor of a layer of another metal, e.g. $\text{Re}$. In this paper
we have not discuss complicated mechanisms involving the combined effects of
electron and nuclear spins such as electron spin rotation induced by nuclear
spin nearby. We believe that it is unlikely that these mechanisms can
produce sufficiently high upper frequency cutoff and sufficient noise level
for a natural $\text{Si}$ with a low concentration of nuclear spins but this
should be also verified experimentally by measuring noise on isotopically
pure $\text{Si}$ substrates. Finally, in our model the $1/f$ dependence of
the noise is due to 'fast' diffusion and the divergent dependence of the
probe magnetic field at the edge of the SQUID loop. Thicker SQUID loops have
different spatial current distribution, this should affect the frequency
dependence of the noise.

We acknowledge fruitful discussions with A. Kitaev, S. Lloyd, J. Martinis,
R. McDermott, Y. Nakamura, V. Schmidt and F. Wellstood. This work was
supported by the National Security Agency (NSA) under Army Research Office
(ARO) contract number W911NF-06-1-0208 and NSF ECS 0608842.

\vspace*{-2mm} 
\bibliography{Flux_ShortL5.bbl}

\end{document}